\documentclass[9pt,twocolumn]{article}
\usepackage{spconf} 
\usepackage{amsmath,amssymb}
 \usepackage{mathrsfs}
\usepackage{dsfont}
\usepackage{tabularx}
\usepackage{graphicx}
\usepackage[font=scriptsize,labelfont=bf]{caption}

\usepackage{appendix}
\usepackage{algorithm}
\usepackage{algpseudocode}
\usepackage{algcompatible}

\DeclareMathOperator{\argmin}{argmin}
\DeclareMathOperator{\argmax}{argmax}

\newcommand{\brac}[1]{\left[#1\right]}
\newcommand{\set}[1]{\left\{#1\right\}}

\newcommand{\paren}[1]{\left(#1\right)}
\newcommand{\kl}[2]{\mathds{KL}\left( #1 \| #2 \right)}

\begin{document}
\allowdisplaybreaks
\setlength{\textfloatsep}{5pt}
\setlength{\abovedisplayskip}{2pt}
\setlength{\belowdisplayskip}{3pt}
\title{Semi-supervised Multi-sensor classification via Consensus-based Multi-View Maximum Entropy Discrimination}
\name{Tianpei Xie$^{ \dagger}$,\qquad Nasser M. Nasrabadi$^{\star}$\qquad and\;\; Alfred O. Hero III$^{ \dagger}$\thanks{Acknowledgement: This research was partially supported by US Army Research Office (ARO) grants W911NF-11-1-0391 and  WA11NF-11-1-103A1.}}
\address{$^{\dagger}$Dept. of Electrical Eng., System,\; University of Michigan, Ann Arbor, MI 48109 \\
          $^{\star}$ U.S. Army Research Lab., 2800 Powder Mill Road, Adelphi, MD, USA\\
$^{\dagger}$ $\{$tianpei, \;hero$\}$@umich.edu, \; $^{\star}$ nasser.m.nasrabadi.civ@mail.mil}
\maketitle
\begin{abstract}
In this paper, we consider multi-sensor classification when there is a large number of unlabeled samples. The problem is formulated under the multi-view learning framework and a Consensus-based Multi-View Maximum Entropy Discrimination (CMV-MED) algorithm is proposed. By iteratively maximizing the stochastic agreement between multiple classifiers on the unlabeled dataset, the algorithm simultaneously learns multiple high accuracy classifiers. We demonstrate that our proposed method can yield improved performance over  previous multi-view learning approaches by comparing performance on three real multi-sensor data sets. 
\end{abstract}

\vspace{-5pt}
\section{Introduction}\vspace{-10pt}
In many applications, e.g., in sensor networks, data is collected from multiple sensors and, given that complementary information is present within different sensors, classification using all sensors is expected to yield higher performance as compared to its single-sensor counterpart \cite{xiong2002multi}. 
Furthermore, as class labeling can be labor intensive, in many situations many training samples may not be labeled. In the machine learning literature, this problem falls under the framework of semi-supervised multi-view learning \cite{zhou2010semi}, since the partially-labeled samples are multi-modal in nature and each modality corresponds to one view of physical event.   

Most methods to multi-sensor or multi-view classification either rely on feature fusion (early fusion) methods, that find an intermediate joint representation of multiple views \cite{lai2000kernel,shon2005learning}, or, on decision fusion (late fusion) methods that combine decisions from multiple models to improve the overall performance \cite{li2002statistical}. 
Unless the features are optimized for multi-view aggregation, there is no guarantee that feature fusion will lead to good classification performance. In this paper, we pursue a different approach that learns an intermediate model, or a \emph{consensus view} to fuse features from different views, and improves simultaneously the performance of each single-view classifier. Moreover, we propose to train a set of \emph{stochastic classifiers}  to handle the large number of unlabeled training samples. 

We follow the principle of the disagreement-based multi-view learning \cite{zhou2010semi, blum1998combining, farquhar2005two, yu2007bayesian,  ganchev2008multi, sindhwani2008rkhs, sun2013multi}. In particular, it is shown in \cite{dasgupta2002pac} that the error rate of each classifier in the multi-view system is bounded above by the rate of disagreement between multiple view-specific classifiers.  In other word, the algorithm that explicitly minimizes the disagreement between multiple view-specific classifiers would learn a set of compatible classifiers with high performance and low sample complexity. In this paper, we propose a Consensus-based Multi-View Maximum Entropy Discrimination (CMV-MED) algorithm that learns a set of classifiers, one for each view, by iteratively maximizing their \emph{stochastic agreement} on the unlabeled training data.  Our method is based on the Maximum Entropy Discrimination (MED) by Jaakkola et al. \cite{jaakkola1999maximum}.  MED is a Bayesian learning approach that generalizes support vector machine (SVM) classifiers and explicitly incorporate the large-margin training \cite{roller2004max} 
into a unified maximum entropy learning framework. 
We show the superior performance of our model over previous multi-view learning approaches by comparing performance on three real multi-sensor data sets. 

This paper is structured as follows: an overview of the MED model is given in Section \ref{Sec: II} and we propose the general model for CMV-MED in Section \ref{Sec: III}. The algorithm for solving CMV-MED is discussed in Section \ref{Sec: IV}. In Section \ref{Sec: V},  experiments on a set of real multi-view data sets are discussed.\vspace{-7pt}

\section{Maximum Entropy Discrimination (MED)}\label{Sec: II}\vspace{-3pt}
We denote the multi-view data set  as $\mathcal{D}_{V}$.  $\mathcal{D}_{V}$ consists of the labeled part $ \set{ (\mathbf{x}_{n}, y_{n}), n \in L   }$  and the unlabeled part  $ \set{ \mathbf{x}_{m}, m \in U   }$, where $L$ and $U$ represent the index set of labeled and unlabeled samples, respectively, and $|L| \ll |U|$. Define the multi-view feature $\mathbf{x}_{n} = [\mathbf{x}_{n}^{1}, \ldots,\mathbf{x}_{n}^{V}], \forall n\in L\cup U$, where $\mathbf{x}_{n}^{i} \in \mathcal{R}^{d_{i}}$ are the features extracted from view $i$ and $V$ is the number of views. Here we consider the \emph{binary classification} task, i.e., $y\in |\mathcal{Y}| = \set{-1, +1}.$ Let $\mathcal{D}^{i}$ be the set of samples collected from the single view $i$. In this section, we focus on the single-view MED on labeled subset $L$.

For a single view $i \in [1,\ldots, V]$,  assume the predictive distribution is a generalized log-linear model, i.e.,
$\log p_{i}(y | \mathbf{x}^{i},\,\mathbf{w}_{i}) \propto \frac{1}{2}y\paren{\mathbf{w}_{i}^{T}\,\Phi_{i}(\mathbf{x}^{i})} \equiv F_{i}(y, \mathbf{x}; \mathbf{w}_{i})$  and  $\Phi_{i}: \mathcal{R}^{d_{i}} \mapsto  \mathcal{R}^{p_{i}} $ is a prescribed feature map defined in view $i$. Define the kernel function $K_{i}: \mathcal{R}^{d_{i}} \times \mathcal{R}^{d_{i}} \mapsto \mathcal{R}$ that satisfies $\langle \boldsymbol{\Phi}_{i}(\mathbf{x}_{n}^{i}),\,\boldsymbol{\Phi}_{i}(\mathbf{x}_{m}^{i})\rangle = K_{i}(\mathbf{x}_{n}, \mathbf{x}_{m}) $, for $\forall \mathbf{x}_{n}^{i}, \mathbf{x}_{m}^{i} \in \mathcal{D}^{i}$ in view $i$ and $F_{i}(y, \mathbf{x}^{i}; \mathbf{w}_{i})$ is the normalized log-likelihood function parameterized by $\mathbf{w}_{i}$ in the kernel space.

Denote the prior distribution of $\mathbf{w}_{i}$ as $p_{0}(\mathbf{w}_{i})$. The goal for Maximum Entropy Discrimination \cite{jaakkola1999maximum} is to learn a post-data (posterior) distribution $ q(\mathbf{w}_{i})$,  by solving an entropic regularized risk minimization problem with the prior on model parameter $\mathbf{w}_{i}$ specified as $p_{0}(\mathbf{w}_{i})$
\begin{small} 
\setlength{\abovedisplayskip}{4pt}
\setlength{\belowdisplayskip}{2pt}
\begin{eqnarray}
&&\min_{q(\mathbf{w}_{i})}  \mathds{KL}\left(q(\mathbf{w}_{i})\| p_{0}(\mathbf{w}_{i})\right)  \nonumber\\[-1pt]
&&+ \sum_{n \in L}\left[1 - \mathds{E}_{q(\mathbf{w}_{i})}\{\Delta F_{i}(y_{n}, \mathbf{x}_{n}^{i}; \mathbf{w}_{i})\} \right]_{+}, \label{eqn: med-class_constrain_1} 
\end{eqnarray}
\end{small}
\hspace{-10pt} where $[s]_{+} = \max\{s, 0\}$. $\mathds{KL}(p \| q)$ is the \emph{Kullback-Leibler divergence} from distribution $p$ to $q$, i.e.,
$\footnotesize{ \kl{q(\mathbf{w}_{i})}{p_{0}(\mathbf{w}_{i})} = \int_{\Theta}q(\mathbf{w}_{i})\log\left( \frac{q(\mathbf{w}_{i})}{p_{0}(\mathbf{w}_{i})}\right) d\mathbf{w}_{i}}$ and 
$\Delta F_{i}(y_{n}, \mathbf{x}_{n}; \mathbf{w}_{i}) \equiv F_{i}(y_{n}, \mathbf{x}^{i}_{n}; \mathbf{w}_{i}) - F_{i}(y \neq y_{n}, \mathbf{x}_{n}^{i}; \mathbf{w}_{i}) =  \log\left(\frac{p(y_{n} | \mathbf{x}_{n}^{i},\,\mathbf{w}_{i})}{p(y\neq y_{n} | \mathbf{x}_{n}^{i},\,\mathbf{w}_{i})} \right) $ is the log-odds classifier.  

The second term in \eqref{eqn: med-class_constrain_1} is a hinge-loss that captures the large-margin principle underlying the MED prediction rule,
\setlength{\abovedisplayskip}{4pt}
\setlength{\belowdisplayskip}{3pt}
\begin{equation*}
y^{*} = \argmax_{y} \mathds{E}_{q(\mathbf{w}_{i})}\left[F(y, \mathbf{x}^{i}; \mathbf{w}_{i})\right]. 
\end{equation*} 

If we use a \emph{Gaussian Process} \cite{rasmussen2006gaussian} as the prior on $\mathbf{w}_{i}$, i.e., $p_{0}(\mathbf{w}_{i}) = \mathcal{N}(\mathbf{w}_{i};\,0, \sigma^{2}I_{p_{i}})$, a kernel SVM is obtained by solving \eqref{eqn: med-class_constrain_1} in its dual formulation.  For multi-view data, it is necessary to learn multiple MEDs simultaneously. For example,  in \cite{jebara2011multitask}, the author applies a joint sparsity prior on $(\mathbf{w}^{1},\ldots,\mathbf{w}^{V})$ to achieve multi-task feature selection. Instead of assuming a joint prior on all multi-view model parameters, we utilize the available unlabeled samples and require the class prediction of multiple models to agree with each other. \vspace{-6pt}

\section{Consensus-based Multi-view MED: a general framework }\label{Sec: III}
Define the \emph{consensus view model} as a parameter-free distribution $q(y |  \mathbf{x}_{n}) \in \mathcal{Q}$ on the unlabeled set $U$, where $\mathbf{x}_{n} = [\mathbf{x}_{n}^{1}, \ldots,\mathbf{x}_{n}^{V}], \forall n\in U$, $\mathcal{Q} \equiv \set{q(x): q(x)\ge 0, \int q(x) dx = 1 }$ and $q(y |  \mathbf{x}_{n}) = \delta\set{y=y_{n}},\; n\in L.$ In each view $i$, a joint post-data distribution is obtained as $q_{i}(y, \mathbf{w}_{i} |  \mathbf{x}) = q(y|  \mathbf{x})q( \mathbf{w}_{i} )$, where  $q(y |  \mathbf{x}) $ is \emph{shared} among all views and the above equality reflects the mean-field approximation.  

The \emph{goal} of  \emph{Consensus-based Multi-view Maximum Entropy Discrimination} (CMV-MED) is to simultaneously learn the joint post-data distributions $q_{i}(y, \mathbf{w}_{i} |  \mathbf{x})= q(y|  \mathbf{x})q( \mathbf{w}_{i} )$, given the priors $p_{i}(y, \mathbf{w}_{i} |  \mathbf{x}^{i}) = p_{i}(y |   \mathbf{w}_{i},  \mathbf{x}^{i})p_{0}(  \mathbf{w}_{i})$  for $\mathbf{x}^{i} \in \mathcal{D}^{i}, \forall i=1,\ldots, V.$ This is accomplished by solving the following optimization problem 
\begin{small}
\setlength{\abovedisplayskip}{3pt}
\setlength{\belowdisplayskip}{3pt}
\begin{flalign}
 \min_{q_{i}(y,\mathbf{w}^{i}| \mathbf{x}_{n}) \in \mathcal{Q}, \atop \forall i=1,\ldots, V, \; n\in L\cup U  }  \sum_{n \in L}\sum_{i=1}^{V}\left[1 - \mathds{E}_{q_{i}(y, \mathbf{w}^{i}|\mathbf{x}_{n})}\{\Delta F_{i}(y, \mathbf{x}_{n}^{i}; \mathbf{w}_{i})\} \right]_{+} \nonumber\\[-2pt]
+ \lambda\sum_{n\in U}\sum_{i=1}^{V}\pi_{i}\mathds{KL}\left(q_{i}(y, \mathbf{w}^{i}|\mathbf{x}_{n})\|  p_{0}(y, \mathbf{w}^{i}|\mathbf{x}_{n}^{i})\right),\label{expr:  model_Bdist_v1}
\end{flalign} \end{small}
\hspace{-5pt}where $\pi_{i} \in \set{\pi_{j}: \sum_{j=1}^{V}\pi_{j} = 1, \quad  \pi_{j}\ge 0, \forall j} $ is a parameter for view $i$ and $\lambda>0$ is regularization parameter. Note that $q_{i}(y, \mathbf{w}_{i} |  \mathbf{x}_{n})= \delta\set{y=y_{n}}q( \mathbf{w}_{i} )$ on the labeled set $L$ and the second term can be further expanded as \vspace{-2pt}
\begin{small}
\begin{flalign}
\setlength{\abovedisplayskip}{1pt}
\setlength{\belowdisplayskip}{1pt}
\hspace{-15pt}\mathds{KL}\left(q_{i}(y, \mathbf{w}^{i}|\mathbf{x}_{n})\|  p_{0}(y, \mathbf{w}^{i}|\mathbf{x}_{n}^{i})\right)
= \kl{q(\mathbf{w}^{i})}{p_{0}(\mathbf{w}^{i})}  \nonumber\\
+ \mathds{E}_{q(\mathbf{w}^{i})}\brac{\kl{q(y|\mathbf{x}_{n})}{p_{i}(y| \mathbf{x}_{n}^{i}, \mathbf{w}^{i})}} 
 \; ,i= 1,\ldots,V. \label{eqn: kl_expand} \vspace{-3pt}
\end{flalign}
\end{small}
\hspace{-2pt}Substituting \eqref{eqn: kl_expand} into \eqref{expr:  model_Bdist_v1}, we have the following\vspace{-3pt}
\begin{small}
\begin{eqnarray}
\setlength{\abovedisplayskip}{1pt}
\setlength{\belowdisplayskip}{2pt}
 \min_{q(y | \mathbf{x}_{n}) \in \mathcal{Q}, n \in U\atop q(\mathbf{w}^{i}),\,\forall i=1,\ldots, V }  \sum_{n \in L}\sum_{i=1}^{V}\left[1 - \mathds{E}_{q(\mathbf{w}^{i})}\{\Delta F_{i}(y_{n}, \mathbf{x}_{n}^{i}; \mathbf{w}_{i})\} \right]_{+} \nonumber\\[-3pt]
+\lambda\sum_{i=1}^{V}\pi_{i}\kl{q(\mathbf{w}^{i})}{p_{0}(\mathbf{w}^{i})} \nonumber\\
+ \lambda\sum_{n\in U}\sum_{i=1}^{V}\pi_{i}\mathds{E}_{q(\mathbf{w}^{i})}\brac{\kl{q(y|\mathbf{x}_{n})}{p_{i}(y| \mathbf{x}_{n}^{i}, \mathbf{w}^{i})}}.
\label{expr:  model_Bdist_v2}  
\end{eqnarray}\end{small} 
\hspace{-3pt}From \eqref{expr:  model_Bdist_v2}, we see that the first and second term learn $V$ view-specific MED models $q(\mathbf{w}^{i}),i=1,..,V,$ simultaneously. 

\begin{figure}[htb] 
\begin{minipage}[b]{1\linewidth}
  \centering
  \centerline{\includegraphics[trim = 10mm 65mm 32mm 60mm, clip, scale = 0.28]{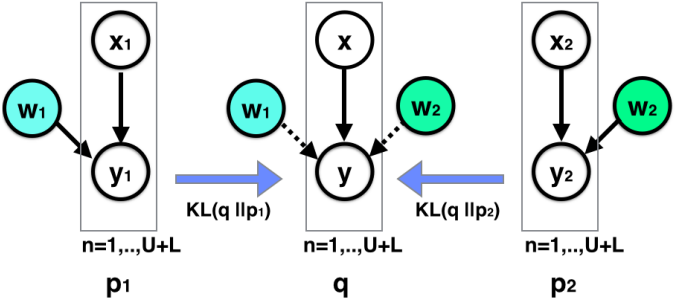}}
\end{minipage}
\caption{\footnotesize{A graphical model representation for consensus-based multi-view learning via information projection.}}
\label{fig: pgm_mv}
\end{figure}

\emph{Our main contribution} is the third term in \eqref{expr:  model_Bdist_v2}, which is referred as the \emph{consensus-based disagreement term} on unlabeled set, since it is zero when view-specific predictive models $p_{i}(y| \mathbf{x}_{n}^{i}, \mathbf{w}^{i})$ all equal, $i=1,...,V$, while it penalizes more when one deviates far from the consensus model $q(y|\mathbf{x})$, which, by construction, is the \emph{center} of these $V$ distributions in the information geometry over the space of probability measures. This center is determined by information projection accomplished by the KL divergence in \eqref{expr:  model_Bdist_v2}. By incorporating this term, we explicitly require all classifiers to make similar class predictions having similar confidence levels on the unlabeled training samples. The benefit for enforcing the consensus-based disagreement is that the proposed model is sensitive in the case when view-specific classifiers with low confidence agree with each other, while it is lenient when all of them are highly confident and agree. Thus the model is reliable in the situation where the initial view-specifc classifiers only have low confidence results due to the limited size of labeled training set. 
Fig. \ref{fig: pgm_mv} is a graphical model representation for the information projection. \vspace{-8pt}

\vspace{-3pt}


\section{Solution via deterministic annealing Expectation Maximization}\label{Sec: IV}
Our solution for CMV-MED in \eqref{expr:  model_Bdist_v2} is based on the \emph{deterministic annealing EM} \cite{sindhwani2006deterministic}. It is described as the following steps: 
\begin{enumerate}
\item Set the regularization parameter $\lambda_{0} = 0$ in \eqref{expr:  model_Bdist_v2} at initialization and train $V$ independent MED classifiers simultaneously to find $q_{0}(\mathbf{w}^{i})$, $i=1,\ldots,V$.  Set the prior distribution $p_{0}(\mathbf{w}^{i}) = \mathcal{N}(\mathbf{w}^{i}: 0, \sigma^{2}I)$ and $\pi_{i} = \frac{1}{V}, \forall i.$ Let $T$ be the maximum number of iterations. \vspace{-3pt}
\item For $t=1,\ldots, T$, do
     \begin{enumerate}
      \item Given the post-data distribution $q_{t-1}(\mathbf{w}^{i}), i=1,\ldots,V$ from MED,  find the consensus view on unlabeled data $U$ via information projection, i.e. \vspace{-3pt}
      \begin{small}
      \begin{eqnarray}
      \hspace{-15pt}q_{t}(y| \mathbf{x}_{n}) \hspace{0.8\linewidth} \nonumber \\[-5pt]
     \hspace{-15pt}  = \argmin_{q} \frac{1}{V}\sum_{i=1}^{V}\mathds{E}_{q(\mathbf{w}^{i})}\brac{\kl{q_{n}(y)}{ p_{i,n}(y|\mathbf{w}^{i}) }}\nonumber\\[-5pt]
      \hspace{-15pt} \Rightarrow  \log q_{t}(y| \mathbf{x}_{n}) =  \frac{1}{V}\sum_{i=1}^{V}\log p_{i,n}(y|\hat{\mathbf{w}}_{t-1}^{i}) - \log Z(\mathbf{x}_{n}), \hspace{-5pt}\nonumber\\[-8pt]
        \forall n\in U, \hspace{0.8\linewidth} \nonumber  \vspace{-3pt}
      \end{eqnarray}       
      \end{small}
  \hspace{-9pt} where $q_{n}(y) \equiv$ $q(y|\mathbf{x}_{n})$, $p_{i,n}(y|\mathbf{w}^{i})\equiv p_{i}(y| \mathbf{x}_{n}^{i},$ $\mathbf{w}^{i})$
   for $n \in U$, $Z(\mathbf{x}_{n})$ is the normalization factor and $\hat{\mathbf{w}}_{t-1}^{i}$ is the mean of the post-data distribution $q_{t-1}(\mathbf{w}^{i}), i=1,\ldots,V$. \vspace{2pt}
   
     \item Given the consensus view $q_{t}(y| \mathbf{x}_{n}), \forall n\in U$, substitute it into \eqref{expr:  model_Bdist_v2} to obtain the following optimization problem
      \begin{small}
    \begin{eqnarray}
      \hspace{-15pt}\min_{q(\mathbf{w}^{i}),\,\forall i=1,\ldots, V }  \sum_{n \in L}\sum_{i=1}^{V}\left[1 - \mathds{E}_{q(\mathbf{w}^{i})}\{\Delta F_{i}(y_{n}, \mathbf{x}_{n}^{i}; \mathbf{w}_{i})\} \right]_{+} \nonumber\\
       \hspace{-15pt} + \lambda_{t}\frac{1}{V}\sum_{n\in U}\sum_{i=1}^{V}\mathds{E}_{q(\mathbf{w}^{i})}\brac{\mathds{E}_{q_{t}(y| \mathbf{x}_{n})}\brac{-\log p_{i}(y| \mathbf{x}_{n}^{i}, \mathbf{w}^{i})}} \nonumber\\
 \hspace{-25pt} +\sum_{i=1}^{V}\pi_{i}\kl{q(\mathbf{w}^{i})}{p_{0}(\mathbf{w}^{i})} \nonumber
      \end{eqnarray}       
      \end{small}
 \hspace{-2pt}For each view $i$,  compute the $q_{t}(\mathbf{w}^{i}| \mathcal{D}^{i},\boldsymbol{\alpha}^{i}) $ with dual parameter $ \boldsymbol{\alpha}^{i} = [\alpha_{1}^{i},\ldots, \alpha_{L}^{i}]^{T}$ by solving the following dual programming problem, i.e.,  
    \begin{small}    
      \begin{eqnarray}
&&\max _{\boldsymbol{\alpha}^{i} }
\mathbf{1}^{T}\boldsymbol{\alpha}^{i}-  \frac{\sigma^{2}}{2}(\boldsymbol{\alpha}^{i})^{T}(\widetilde{\mathbf{K}}_{i}\odot \mathbf{y}\,\mathbf{y}^{T})\boldsymbol{\alpha}^{i} \label{eqn: dual_form} \\
&& \text{s.t. } \mathbf{0} \preceq \boldsymbol{\alpha}^{i} \preceq \mathbf{1}, \nonumber
\end{eqnarray}
  \end{small}
 \hspace{-5pt}where $\mathbf{1} = [1,\ldots, 1]^{T}$ and $\odot$ is piece-wise product.  In \eqref{eqn: dual_form}, \emph{a new kernel} $\widetilde{\mathbf{K}}_{i}$ is computed via
 \begin{small}
  \begin{eqnarray}
  &&\widetilde{\mathbf{K}}_{i} = \mathbf{K}_{L,i} \nonumber\\
  &&-\lambda_{t}\, (\mathbf{k}_{UL}^{i})^{T} \left[ 1/\sigma^{2}\mathbf{M}_{i}^{-1} +  \lambda_{t}\mathbf{K}_{U,i}\right]^{-1}\mathbf{k}_{UL}^{i} \\
  && \equiv [\langle \widetilde{\Phi}_{i}(\mathbf{x}_{n}^{i})\; , \; \widetilde{\Phi}_{i}(\mathbf{x}_{m}^{i}) \rangle]_{n,m\in L},
  \end{eqnarray}
  \end{small}    
 \hspace{-15pt}where $\mathbf{K}_{L,i} = [K_{i}(\mathbf{x}_{n}^{i}, \mathbf{x}_{m}^{i})]_{n,m\in L}$, $\mathbf{K}_{U,i} = [K_{i}(\mathbf{x}_{n}^{i}, \mathbf{x}_{m}^{i})]_{n,m\in U}$ and $\mathbf{k}_{UL}^{i} = [K_{i}(\mathbf{x}_{n}^{i}, \mathbf{x}_{m}^{i})]_{n \in U,m\in L}$. $\mathbf{M}_{i} = diag\set{\nu_{1}, \ldots, \nu_{U}} \in \mathcal{R}^{|U| \times |U|}$, with $\nu_{n} \equiv  \mathds{E}_{q_{t}(y| \mathbf{x}_{n})}\brac{-\nabla_{\mathbf{w}^{i}}^{2} \log p_{i}(y| \mathbf{x}_{n}^{i}, \hat{\mathbf{w}}_{t-1}^{i})  }, n\in U$. \\[-5pt]
  
  Then the post-data distribution $q_{t}(\mathbf{w}^{i}| \mathcal{D}^{i},\boldsymbol{\alpha}^{i}) = \mathcal{N}(\hat{\mathbf{w}}_{t}^{i}, \mathbf{H}_{i}) $, where the mean is given by $\hat{\mathbf{w}}_{t}^{i} = \sum_{m=1}^{L}y_{m}\alpha_{m}^{i}\widetilde{\Phi}_{i}(\mathbf{x}_{n}^{i})$. The covariance matrix $\mathbf{H}_{i} = \paren{\sigma^{2}\,I + \boldsymbol{\Phi}_{i}(\mathbf{X}_{U})^{T}\,\mathbf{M}_{i}\,  \boldsymbol{\Phi}_{i}(\mathbf{X}_{U})}$ with $\boldsymbol{\Phi}(\mathbf{X}_{U}) \equiv [\Phi_{i}(\mathbf{x}_{1}^{i}),\ldots, \Phi_{i}(\mathbf{x}_{U}^{i})]^{T} \in \mathcal{R}^{|U| \times p_{i}}$. 
  \item Set $\lambda_{t} = 1 - e^{-0.5t} \rightarrow 1$ as $t$ increases. 
  \item $t \leftarrow t+1.$
     \end{enumerate} 
\item Finally, make prediction based on consensus view
\begin{small} 
\setlength{\abovedisplayskip}{2pt}
\setlength{\belowdisplayskip}{2pt}
\begin{equation*}
y^{*} = \argmax_{\hat{y}} \sum_{1\le i \le V}\mathds{E}_{q(y, \mathbf{w}_{i})}\left[\delta\set{y= \hat{y}}F(y, \mathbf{x}^{i}; \mathbf{w}_{i})\right]. 
\end{equation*} 
\end{small}   
\end{enumerate}
Note that the Step 2(b) can be performed in parallel, as  it does not rely on information from other views. 

\vspace{-8pt}
\section{Experiments}\label{Sec: V}\vspace{-6pt}
We compare the proposed CMV-MED model with the SVM-2K model proposed by Farquhar et al. \cite{farquhar2005two}, the MV-MED model by Sun et al. \cite{sun2013multi} as well as the conventional MED for each view on several real multi-view data sets. 
In the following experiments, we focus on two-view learning, i.e. $V=2$ and use the Gaussian Kernel function $K_{i}(\mathbf{x}^{i}_{n}, \mathbf{x}^{i}_{m}) = \exp(c\,\|\mathbf{x}^{i}_{n} -  \mathbf{x}^{i}_{m}\|^{2}), i=1,2$. For all MED-based methods, a Gaussian Process prior $p_{0}(\mathbf{w}^{i}) = \mathcal{N}(\mathbf{0}, \sigma_{i}^{2}I)$ is assigned for view $i=1,2.$ The view parameter $\pi_{1} = \pi_{2} = \frac{1}{2}$. All other parameters for each model are obtained by 5-fold-cross-validation.  All the experiments are repeated for $20$ times, with randomly chosen $L$ and $U$.
\begin{table*}[tb]
 \footnotesize 
 \centering
\begin{tabular*}{0.91\textwidth}{|m{107pt}<{\centering}|m{50pt}<{\centering}|m{50pt}<{\centering}|m{60pt}<{\centering}|m{60pt}<{\centering}|m{60pt}<{\centering}|}
\hline 
\multicolumn{6}{|c|}{Classification Accuracy ($\%$) mean $\pm$ standard error} \\ 
\hline
Dataset.& \multicolumn{2}{c|}{ MED (single views) } & SVM-2K  &  MV-MED    & CMV-MED \\ 
\hline 
\textbf{ARL Footstep} (Sensor 1,2, $|L|=50$) & $71.1\pm 5.3$ & $62.3 \pm 10.2$  & $73.3 \pm 5.2$ & $75.6 \pm 6.5$  & $\mathbf{85.5} \pm 6.1$ \\ 
\hline 
\textbf{WebKB4} ($|L|=15 $)&  $76.6 \pm 10.2$  & $77.1 \pm 10.1 $ & $79.0 \pm 10.0$ & $77.9 \pm 8.7 $  & $\mathbf{91.7} \pm 5.8$ \\ 
\hline 
\textbf{Internet Ads} ($|L|= 50$)& $87.3 \pm 0.9$ & $86.2 \pm 1.4$  & $82.5 \pm 4.3$ &  $88.8 \pm 2.3$ & $\mathbf{92.7} \pm 0.7 $\\ 
\hline 
\end{tabular*} \vspace{-5pt}
\caption{\footnotesize Classification accuracy with different data set, with the best performance shown in \textbf{bold}.} 
\label{tab: class_accuracy}
\end{table*}
\vspace{-9pt}
\begin{figure*}[tb] 
\begin{minipage}[b]{0.28\linewidth}
  \centering
  \centerline{\includegraphics[scale = 0.4, trim = 2mm 2mm 10mm 2mm, clip]{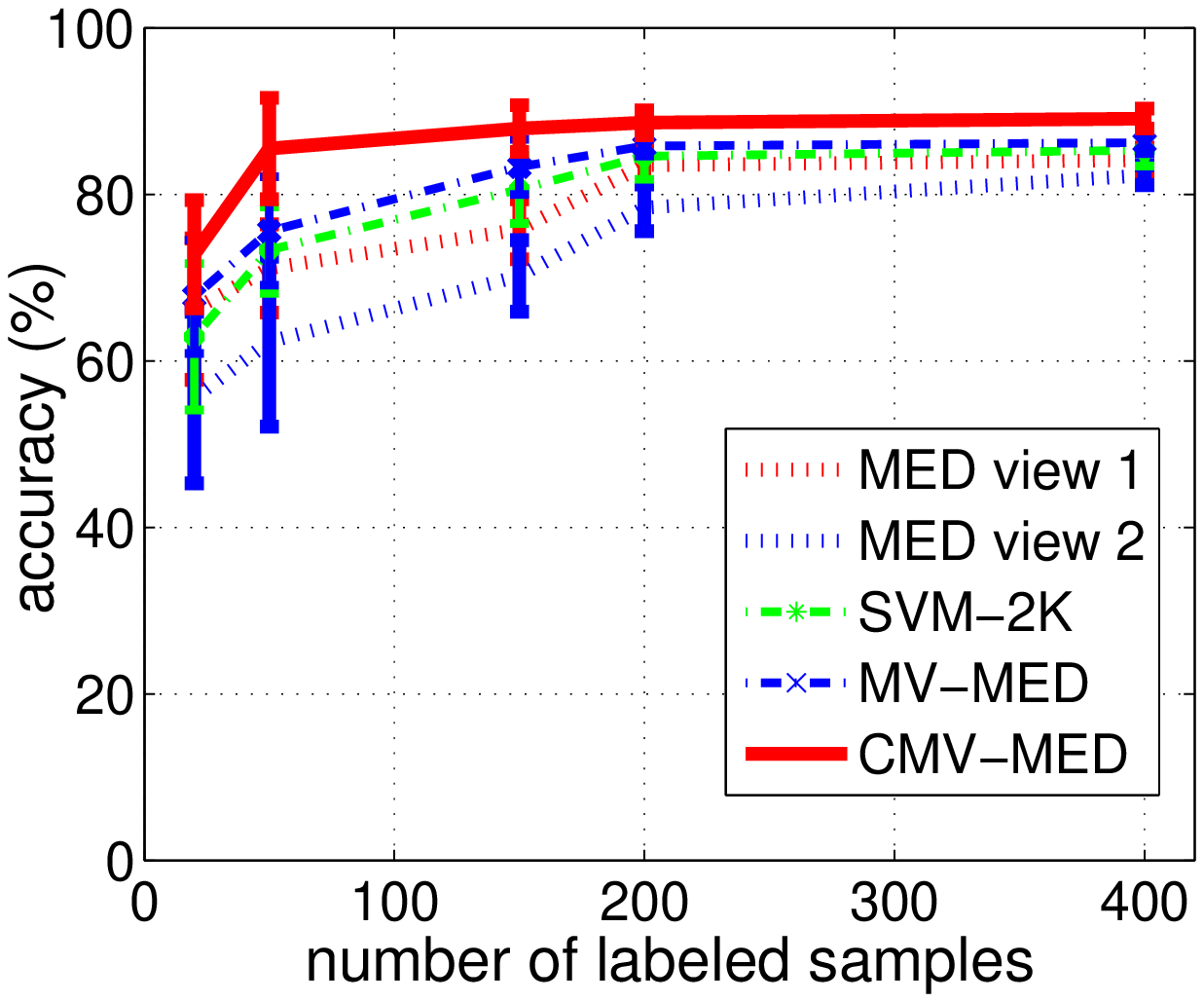}}
  \vspace{-7pt}
  \centerline{\footnotesize{(a)}}
\end{minipage}
\hfill
\begin{minipage}[b]{0.28\linewidth}
  \centering
  \centerline{\includegraphics[scale = 0.38]{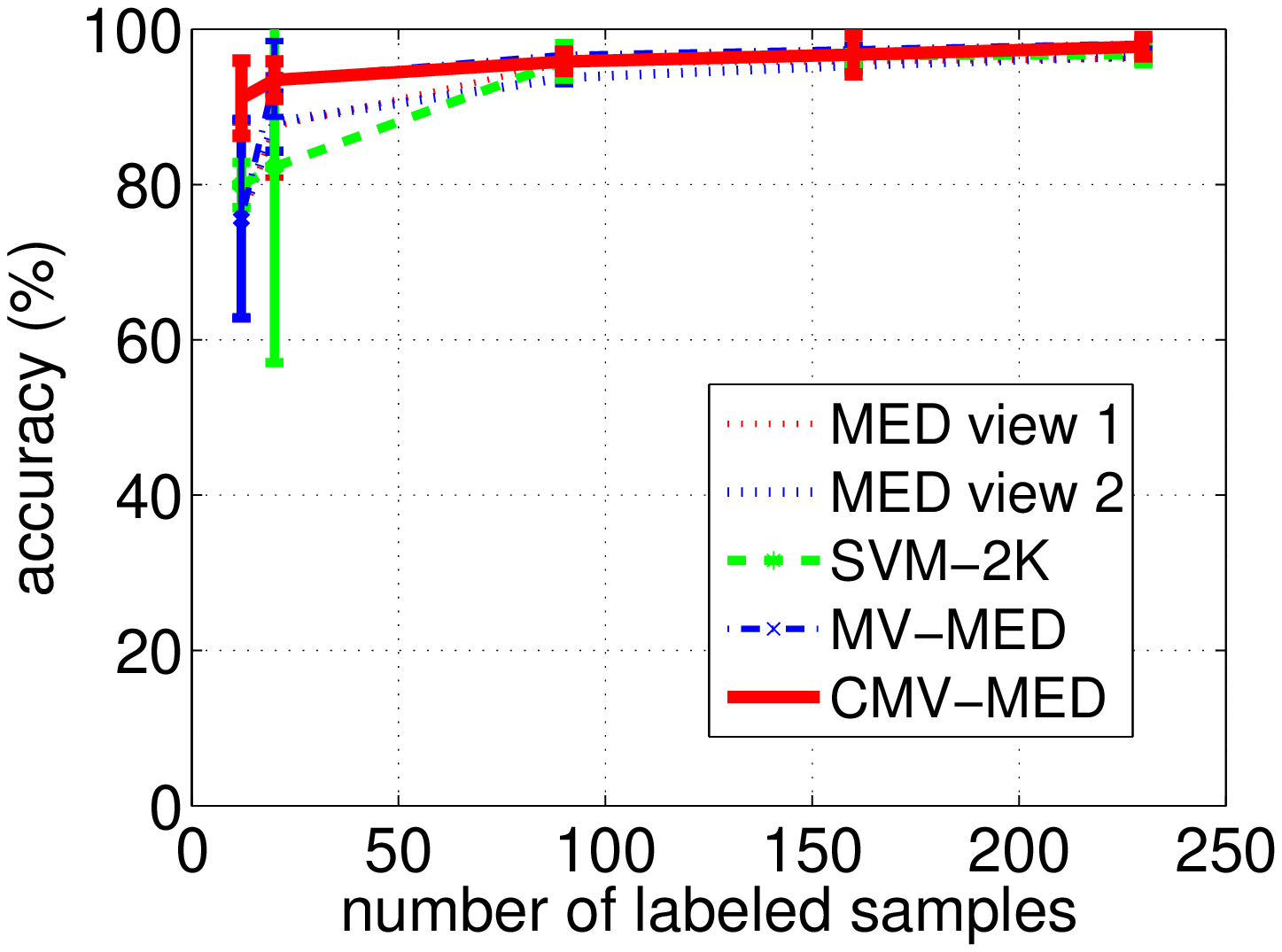}}
  \vspace{-7pt}
 \centerline{\footnotesize{(b) }}
\end{minipage}
\hfill
\begin{minipage}[b]{0.3\linewidth}
  \centering
  \centerline{\includegraphics[scale = 0.38]{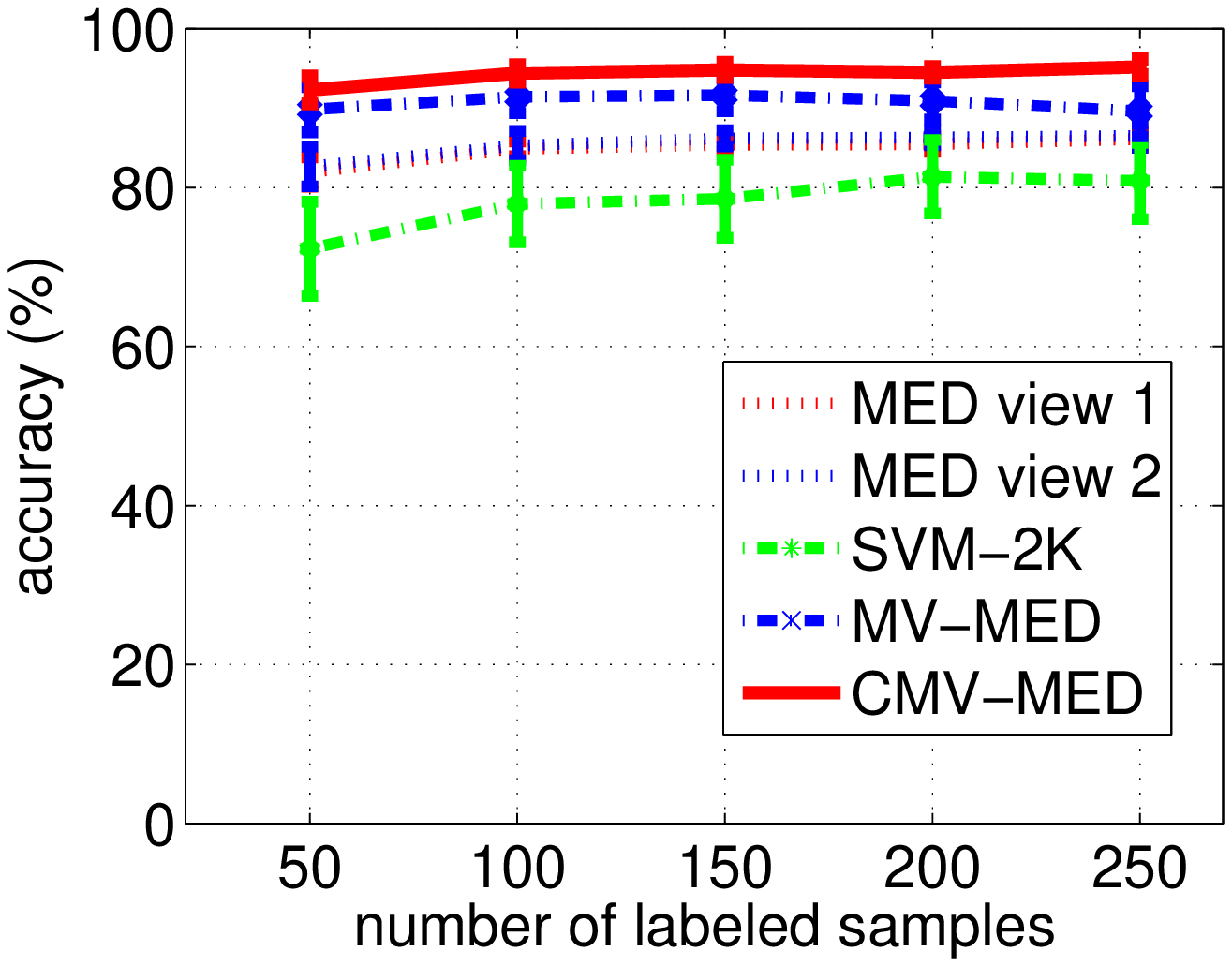}}
  \vspace{-7pt}
 \centerline{\footnotesize{(c) }}
\end{minipage}\vspace{-8pt}
\caption{\footnotesize{The classification accuracy vs. the size of labeled set for (a) \textbf{ARL-Footstep} data set, (b) \textbf{WebKB4} data set and (c) \textbf{Internet Ads} data set. The proposed CMV-MED outperforms MV-MED, SVM-2K and two single-view MEDs (view 1 and 2) and it has good stability when the number of labeled samples is small.}}\vspace{-10pt}
\label{fig: unlabel_vary}
\end{figure*}\vspace{-3pt}
\subsection{Footstep Classification}\vspace{-5pt}
We test on \textbf{ARL-Footstep} \cite{damarla2011detection,nguyen2011robust} data, which is a multi-sensor data set that contains acoustic signals collected by four well-synchronized sensors (labeled as Sensor 1,2,3,4) in a natural environment. The task is to discriminate between human footsteps and human-leading animal footsteps. We only use Sensor $1,2$ in our experiment.  It involves $840$ segments from human subjects and $660$ segments from human-animal subjects. We choose $600$ segments from each class as the training set with $|L| = 50$, and the rest is designated as the test set. A $200$-dimensional mel-frequency cepstral coefficients (MFCCs) vector is computed from the corresponding segments in all the views, with normalization as in \cite{nguyen2011robust}. 

In Table \ref{tab: class_accuracy}, we see that our CMV-MED outperforms both SVM-2K and MV-MED, and it improves over the single-view MED. This is likely because our method utilizes the confidence as well as decision as a disagreement measure, 
In \textbf{ARL-Footstep} data, since the signal is contaminated by background noise, the original MED on two single views does not perform well, and both the decision regularization and margin regularization are not as reliable as the confidence regularization implemented by CMV-MED. 

Fig. \ref{fig: unlabel_vary}(a) shows the accuracy and the standard deviation for the four methods as the size of the labeled set increases. As more ground truth labels are used, the performances of all training methods increases, while CMV-MED shows its superior performance consistently. 

\vspace{-10pt}
\subsection{Web-Page Classification}\vspace{-5pt}
The \textbf{WebKB4} \cite{craven2000learning} data set is widely-used in multi-view learning literature \cite{blum1998combining,sindhwani2008rkhs}. It consists of $1051$ two-view web pages collected from computer science department web sites at four universities. There are $230$ course pages and $821$ non-course pages. The two natural views are words in a web page and words appearing in the links pointing to that page. We follow the preprocessing step in \cite{sindhwani2008rkhs}, and extract a $3000$-dimensional feature vector via the bag-of-words representation in the page view and a $1840$-dimensional feature vector in the link view. Then we compute the term frequency-inverse document frequency weights (TF-IDF) features from the document word matrix. The feature vector is length normalized. 

In Table \ref{tab: class_accuracy}, we see that our CMV-MED has significantly better performance as compared to SVM-2K and MV-MED, when the labeled set is small, i.e., $|L| = 15$. Also, according to Fig. \ref{fig: unlabel_vary}(b), when more labeled samples are included, all four methods have similarly good performance, even for the single-view MED. The CMV-MED performs better with a few labeled samples because its stability relies on a good estimate of confidence on the unlabeled training samples, which is less affected by the amount of the labeled training samples.

\vspace{-12pt}
\subsection{Internet Advertisement Classification}\vspace{-5pt}
The \textbf{Internet Ads} \cite{kushmerick1999learning} data set consists of $3279$ instances including $458$ ads images and $2820$ non-ads images. The first view describes the image itself, i.e., words in images' URL and caption, while the other view contains all other features, i.e., words from URLs of pages that contain the image and pages which the image points to.  For each view, we extract the bag-of-words representations, which results in a $587-$dimensional vector in view 1 and a $967-$dimension vector in view 2. We set the size of training set as $600$ and $|L|= 50$.

From Table \ref{tab: class_accuracy} and Fig. \ref{fig: unlabel_vary}(c) , we see that our CMV-MED still performs better than SVM-2K, MV-MED and single-view MED. It is seen that CMV-MED is more stable as the size of the labeled training set increases, while SVM-2K has much worse stability performance. \vspace{-12pt}

\section{Conclusion}\vspace{-8pt}
In this paper, we propose a consensus-based multi-view maximum entropy learning model that incorporates large-margin classification and Bayesian learning when a large amount of unlabeled samples from multiple sources are available. 
The experimental results on three different real data sets show the superiority of the proposed CMV-MED over other multi-view large-margin classification methods in terms of classification accuracy, especially when the number of labeled samples is small compared to the unlabeled ones. 
\bibliographystyle{IEEEbib}
\bibliography{tianpei_ICASSP15.bib}
\end{document}